\documentclass[12pt]{article}
\title{
\vspace{-3mm}
\rightline{\small IFUP-TH 2002/13}
\vspace{8mm}
\bf Higgs-inspired corrections to the
RG flow in the finite-temperature 3D Georgi-Glashow model and its SU(N)-generalization}
\author{Dmitri Antonov \thanks{
E-mail: {\tt antonov@df.unipi.it}} \thanks{
Permanent address: ITEP, B. Cheremushkinskaya 25, RU-117 218 Moscow, Russia.}
\\
{\it INFN-Sezione di Pisa, Universit\'a degli studi di Pisa,}\\
{\it Dipartimento di Fisica, Via Buonarroti, 2 - Ed. B -
I-56127 Pisa, Italy}}
\date{}
\begin{document}
\maketitle
\vspace{1mm}
\centerline{\bf {Abstract}}
\vspace{3mm}
\noindent
The Berezinsky-Kosterlitz-Thouless (BKT) RG flow in the ensemble of monopoles existing in the
finite-temperature (2+1)D Georgi-Glashow model is explored
in the regime when the Higgs field is not infinitely heavy, but its mass is rather
of the same order of magnitude as the mass of the W-boson. The corrections
to the standard RG flow are derived to the leading order in the inverse mass
of the Higgs boson. According to the obtained RG equations,
the scaling of the free-energy density in the critical region and the
value of the critical temperature of the
phase transition are found to be unaffected by the finiteness of the
Higgs-boson mass. The evolution of the Higgs mass itself is also investigated and
shown to be rather weak, that enables one to treat this parameter as a constant.
The same analysis is further performed in the $SU(N)$-case at $N>2$, where the RG
invariance is demonstrated to hold only approximately, in a certain sense.
Modulo this approximation, the critical behaviour of the $SU(N)$-model
turns out to be identical to that of the $SU(2)$-one.

\vspace{10mm}

\section{Introduction. The model.}

The (2+1)-dimensional Georgi-Glashow model is known as
an example of the theory allowing for an analytic description
of confinement~\cite{1}. Besides the investigations performed in the compact-QED limit
of this model (where the Higgs boson is infinitely heavy), there have recently
also been done several investigations of the Georgi-Glashow model beyond that limit~\cite{nd, mpla}. The analysis
of ref.~\cite{nd} has been carried out in the Bogomolny-Prasad-Sommerfield (BPS) limit~\cite{bps}, that is the limit
where the Higgs field is much lighter than the W-boson.
In that paper, the theory has been investigated both at zero and
nonzero temperatures. In ref.~\cite{mpla}, the zero-temperature results of ref.~\cite{nd}
have been generalized to the case intermediate between the BPS- and the compact-QED limits
of the model, when the mass of the Higgs boson is of the same order of magnitude as the
mass of the W-boson. The aim of the present letter is to generalize to this case also
the finite-temperature results of ref.~\cite{nd}. In this way, we shall address the corrections to the RG
behaviour of the monopole ensemble, existing in the
model, which stem from the finiteness of the Higgs-boson mass. This will be
done in the next Section. In Section~3, these results will be generalized to the case of the
$SU(N)$ Georgi-Glashow model. The main results of the letter will finally be summarized in the
Conclusions.

The Euclidean action of the (2+1)D Georgi-Glashow
model reads~\cite{1}

\begin{equation}
\label{GG}
S=\int d^3x\left[\frac{1}{4g^2}\left(F_{\mu\nu}^a\right)^2+
\frac12\left(D_\mu\Phi^a\right)^2+\frac{\lambda}{4}\left(
\left(\Phi^a\right)^2-\eta^2\right)^2\right].
\end{equation}
Here, the Higgs field $\Phi^a$ transforms by the adjoint representation,
$D_\mu\Phi^a\equiv\partial_\mu\Phi^a+\varepsilon^{abc}A_\mu^b
\Phi^c$. Next, $\lambda$ is the Higgs coupling constant of dimensionality [mass],
$\eta$ is the Higgs v.e.v. of dimensionality $[{\rm mass}]^{1/2}$, and
$g$ is the electric coupling constant of the same dimensionality.
At the one-loop level, the partition function describing the sector of the theory~(\ref{GG})
which contains dual photons and Higgs bosons reads~\cite{dietz}

\begin{equation}
\label{pf}
{\cal Z}=1+\sum\limits_{N=1}^{\infty}\frac{\zeta^N}{N!}\left(\prod\limits_{i=1}^{N}\int d^3z_i
\sum\limits_{q_i=\pm 1}^{}\right)\exp\left[-\frac{g_m^2}{8\pi}
\sum\limits_{{a,b=1\atop a\ne b}}^{N}\left(\frac{q_aq_b}{|{\bf z}_a-{\bf z}_b|}-
\frac{{\rm e}^{-m_H|{\bf z}_a-{\bf z}_b|}}{|{\bf z}_a-{\bf z}_b|}\right)\right]\equiv
\int {\cal D}\chi{\cal D}\psi {\rm e}^{-S},
\end{equation}
where

\begin{equation}
\label{1}
S=\int d^3x\left[\frac12(\partial_\mu\chi)^2+\frac12(\partial_\mu\psi)^2
+\frac{m_H^2}{2}\psi^2-2\zeta{\rm e}^{g_m\psi}\cos(g_m\chi)\right]
\equiv\int d^3x{\cal L}[\chi,\psi|g_m,\zeta].
\end{equation}
The partition function~(\ref{pf}) describes the grand canonical ensemble of monopoles with the
account for their Higgs-mediated interaction.
In eqs.~(\ref{pf}) and~(\ref{1}), $\chi$ is the dual-photon field, and the field $\psi$ accounts for the Higgs field,
whose mass reads $m_H=\eta\sqrt{2\lambda}$. Note that from eq.~(\ref{pf}) it is straightforward to deduce that when $m_H$
formally tends to infinity, one arrives at the conventional sine-Gordon theory of the
dual-photon field~\cite{1} describing the compact-QED limit of the model.
Next, in the above equations, $g_m$ stands for the magnetic coupling constant
related to the electric one as $g_mg=4\pi$, and
the monopole fugacity $\zeta$ has the form:

\begin{equation}
\label{zeta}
\zeta=\frac{m_W^{7/2}}{g}\delta\left(\frac{\lambda}{g^2}\right)
{\rm e}^{-4\pi m_W\epsilon/g^2}.
\end{equation}
In this formula, $m_W=g\eta$ is the W-boson mass,
and $\epsilon=\epsilon(\lambda/g^2)$ is a certain slowly
varying function, $\epsilon\ge 1$, $\epsilon(0)=1$~\cite{bps},
$\epsilon(\infty)\simeq 1.787$~\cite{kirk}.
As far as the function $\delta$ is concerned,
it is determined by the loop corrections. In what follows, we shall work
in the standard weak-coupling regime $g^2\ll m_W$, which parallels the requirement
that $\eta$ should be large enough to ensure the spontaneous symmetry breaking from
$SU(2)$ to $U(1)$.

It is worth emphasizing that the phase transition to be explored in this letter is
the finite-temperature BKT phase transition~\cite{bkt} (for a review see e.g.~\cite{rev}) in the monopole ensemble, rather than the real
deconfining phase transition in the model~(\ref{GG}). The latter has been recently shown~\cite{W} to be
associated with the deconfinement of W-bosons, and those are not described by the action~(\ref{1}).
As far as the influence of the Higgs field to the {\it deconfining} phase transition is concerned, it has been addressed in
ref.~\cite{PLB}.

At finite temperature $T\equiv1/\beta$, one should supply the fields $\chi$ and $\psi$
with the periodic boundary conditions in the temporal direction, with the period equal to
$\beta$. Because of that, the lines of the magnetic field emitted by a monopole cannot cross
the boundary of the one-period region and consequently, at the distances larger than $\beta$,
should go almost parallel to this boundary, approaching it.
Therefore, monopoles separated by such distances
interact via the 2D Coulomb potential, rather than the 3D one. Since the average distance
between monopoles in the plasma is of the order of $\zeta^{-1/3}$, we see that at $T\ge\zeta^{1/3}$,
the monopole ensemble becomes two-dimensional. Owing to the fact that $\zeta$ is exponentially
small in the weak-coupling regime under discussion, the idea of dimensional reduction is perfectly applicable
at the temperatures of the order of the critical temperature of the BKT
phase transition in the monopole plasma, which is equal to $g^2/2\pi$~\cite{nk}.
The factor $\beta$ at the
action of the dimensionally-reduced theory, $S_{{\rm d.-r.}}=\beta\int d^2x{\cal L}[\chi,\psi|g_m,\zeta]$,
can be removed [and this action can be cast to the original form of eq.~(\ref{1}) with the substitution $d^3x\to d^2x$]
by the obvious rescaling:
$S_{{\rm d.-r.}}=\int d^2x{\cal L}\left[\chi^{\rm new},\psi^{\rm new}|\sqrt{K},\xi\right]$.
Here, $K\equiv g_m^2T$, $\xi\equiv\beta\zeta$,
$\chi^{\rm new}=\sqrt{\beta}\chi$, $\psi^{\rm new}=\sqrt{\beta}\psi$, and in what follows
we shall denote for brevity $\chi^{\rm new}$ and $\psi^{\rm new}$ simply as $\chi$ and $\psi$, respectively.

As it has already been mentioned, the RG analysis of the BKT phase transition in the BPS limit, $m_H\ll m_W$,
has been performed in ref.~\cite{nd}. Since the characteristic momenta of the field $\psi$ are of the
order of $m_H$, in that limit one may not be interested in the renormalization of the field $\psi$. Instead, it is
possible to integrate this field out in the original 3D theory at the very beginning and then to explore the RG behaviour of the resulting
action of the dual-photon field at finite temperature. In the present letter, we shall explore the RG
behaviour of the model in the regime intermediate between the BPS- and the compact-QED limits, $m_H\sim m_W$.
Clearly, in such a case one should take into account the renormalization of the field $\psi$. As well as in ref.~\cite{nd},
there will be derived the RG equation describing the evolution of $m_H$. However, due to the largeness of $m_H$ with respect
to the characteristic momentum scale we are working at (since $m_H$ is of the order of the UV cutoff $m_W$), this evolution will
be shown to be rather weak.

\section{RG analysis in the SU(2)-case.}
In what follows, we shall adapt the usual RG strategy~\cite{kogut} based on the integration over the
high-frequency modes. Note that this procedure will be applied to {\it all} the fields, i.e., not only to $\chi$, but also to $\psi$.
Splitting the momenta into two ranges, $0<p<\Lambda'$ and $\Lambda'<p<\Lambda$, one can define the high-frequency
modes as $h=\chi_\Lambda-\chi_{\Lambda'}$, $\phi=\psi_\Lambda-\psi_{\Lambda'}$, where ${\cal O}_{\Lambda'}({\bf x})=
\int\limits_{0<p<\Lambda'}^{}\frac{d^2p}{(2\pi)^2}{\rm e}^{i{\bf p}{\bf x}}{\cal O}({\bf p})$ and
consequently, e.g., $h({\bf x})=\int\limits_{\Lambda'<p<\Lambda}^{}\frac{d^2p}{(2\pi)^2}{\rm e}^{i{\bf p}{\bf x}}\chi({\bf p})$.
The partition function,

$${\cal Z}_\Lambda=\int\limits_{0<p<\Lambda}^{}{\cal D}\chi({\bf p}){\cal D}\psi({\bf p})
\exp\left\{-S_{\rm d.-r.}\left[\chi_\Lambda,\psi_\Lambda\right]\right\},$$
can be rewritten as follows:

$$
{\cal Z}_\Lambda=\int\limits_{0<p<\Lambda'}^{}{\cal D}\chi({\bf p}){\cal D}\psi({\bf p})
\exp\left[\frac12\int d^2x\chi_{\Lambda'}\partial^2\chi_{\Lambda'}+\frac12
\int d^2x\psi_{\Lambda'}\left(\partial^2-m_H^2\right)\psi_{\Lambda'}\right]{\cal Z}',
$$
where

$$
{\cal Z}'=\int\limits_{\Lambda'<p<\Lambda}^{}{\cal D}\chi({\bf p}){\cal D}\psi({\bf p})
\exp\left[\frac12\int d^2xh\partial^2h+\frac12\int d^2x\phi\left(\partial^2-m_H^2\right)\phi+\right.$$

$$\left.+2\xi{\rm e}^{\sqrt{K}\left(\psi_{\Lambda'}+\phi\right)}\cos\left(\sqrt{K}\left(\chi_{\Lambda'}+h\right)
\right)\right].$$
Owing to the exponential smallness of the fugacity, ${\cal Z}'$ can further be expanded as

$${\cal Z}'\simeq 1+2\xi\int d^2x\left<{\rm e}^{\sqrt{K}\left(\psi_{\Lambda'}+\phi\right)}\right>_{\phi}
\left<\cos\left(\sqrt{K}\left(\chi_{\Lambda'}+h\right)\right)\right>_h
+2\xi^2\int d^2x d^2y\times$$

$$\times\left[
\left<{\rm e}^{\sqrt{K}\left(\psi_{\Lambda'}({\bf x})+\phi({\bf x})\right)}{\rm e}^{\sqrt{K}\left(\psi_{\Lambda'}({\bf y})+\phi({\bf y})\right)}
\right>_{\phi}\left<\cos\left(\sqrt{K}\left(\chi_{\Lambda'}({\bf x})+h({\bf x})\right)\right)
\cos\left(\sqrt{K}\left(\chi_{\Lambda'}({\bf y})+h({\bf y})\right)\right)\right>_h-\right.$$

$$\left.-\left<{\rm e}^{\sqrt{K}\left(\psi_{\Lambda'}({\bf x})+\phi({\bf x})\right)}\right>_{\phi}
\left<{\rm e}^{\sqrt{K}\left(\psi_{\Lambda'}({\bf y})+\phi({\bf y})\right)}
\right>_{\phi}\left<\cos\left(\sqrt{K}\left(\chi_{\Lambda'}({\bf x})+h({\bf x})\right)\right)\right>_h
\left<\cos\left(\sqrt{K}\left(\chi_{\Lambda'}({\bf y})+h({\bf y})\right)\right)\right>_h\right],$$
where

$$
\left<{\cal O}\right>_h\equiv\frac{\int\limits_{\Lambda'<p<\Lambda}^{}{\cal D}\chi({\bf p})
\exp\left(\frac12\int d^2xh\partial^2h\right){\cal O}}{\int\limits_{\Lambda'<p<\Lambda}^{}{\cal D}\chi({\bf p})
\exp\left(\frac12\int d^2xh\partial^2h\right)},~
\left<{\cal O}\right>_{\phi}\equiv\frac{\int\limits_{\Lambda'<p<\Lambda}^{}{\cal D}\psi({\bf p})
\exp\left[\frac12\int d^2x\phi\left(\partial^2-m_H^2\right)\phi\right]{\cal O}}
{\int\limits_{\Lambda'<p<\Lambda}^{}{\cal D}\psi({\bf p})
\exp\left[\frac12\int d^2x\phi\left(\partial^2-m_H^2\right)\phi\right]}.$$
Carrying out the averages we arrive at the following expression for ${\cal Z}'$:

$${\cal Z}'\simeq 1+2\xi A(0)B(0)\int d^2x{\rm e}^{\sqrt{K}\psi_{\Lambda'}}\cos\left(\sqrt{K}\chi_{\Lambda'}\right)+
\left(\xi A(0)B(0)\right)^2\int d^2xd^2y\times$$

\begin{equation}
\label{z}
\times{\rm e}^{\sqrt{K}\left(\psi_{\Lambda'}({\bf x})+\psi_{\Lambda'}({\bf y})\right)}
\sum\limits_{k=\pm 1}^{}
\left[A^{2k}({\bf x}-{\bf y})B^2({\bf x}-{\bf y})-1\right]
\cos\left[\sqrt{K}\left(\chi_{\Lambda'}({\bf x})+k\chi_{\Lambda'}({\bf y})\right)\right],
\end{equation}
where

$$A({\bf x})\equiv{\rm e}^{-KG_h({\bf x})/2},~
B({\bf x})\equiv{\rm e}^{KG_{\phi}({\bf x})/2},~
G_h({\bf x})=\int\limits_{\Lambda'<p<\Lambda}^{}\frac{d^2p}{(2\pi)^2}
\frac{{\rm e}^{i{\bf p}{\bf x}}}{p^2},~
G_{\phi}({\bf x})=\int\limits_{\Lambda'<p<\Lambda}^{}\frac{d^2p}{(2\pi)^2}
\frac{{\rm e}^{i{\bf p}{\bf x}}}{p^2+m_H^2}.$$
Since in what follows we shall take $\Lambda'=\Lambda-d\Lambda$,
the factors $\left[A^{2k}({\bf x}-{\bf y})B^2({\bf x}-{\bf y})-1\right]$, $k=\pm 1$, are small. Owing to this fact, it is convenient
to introduce the coordinates ${\bf r}\equiv{\bf x}-{\bf y}$ and ${\bf R}\equiv\frac12({\bf x}+{\bf y})$, and
Taylor expand eq.~(\ref{z}) in powers of ${\bf r}$. Clearly, this expansion should be performed up to the
induced-interaction term $\sim\xi^2\int d^2R{\rm e}^{2\sqrt{K}\psi_{\Lambda'}({\bf R})}\cos\left(2\sqrt{K}\chi_{\Lambda'}({\bf R})\right)$, that should
already be disregarded. As a result, we obtain the following expression for ${\cal Z}_{\Lambda}$~\footnote{For the
sake of uniformity, we replace $d^2R$ by $d^2x$.}:

$$
{\cal Z}_\Lambda=\int\limits_{0<p<\Lambda'}^{}{\cal D}\chi({\bf p}){\cal D}\psi({\bf p})\exp\left\{
\frac12\int d^2x\psi_{\Lambda'}\left(\partial^2-m_H^2\right)\psi_{\Lambda'}+a_3(\xi A(0)B(0))^2\int d^2x
{\rm e}^{2\sqrt{K}\psi_{\Lambda'}}-\right.$$

\begin{equation}
\label{ren}
\left.-\frac12\int d^2x\left[1+a_2(\xi A(0)B(0))^2\frac{K}{2}{\rm e}^{2\sqrt{K}\psi_{\Lambda'}}\right]
\left(\partial_\mu\chi_{\Lambda'}\right)^2+2\xi A(0)B(0)\int d^2x{\rm e}^{\sqrt{K}\psi_{\Lambda'}}\cos\left(\sqrt{K}\chi_{\Lambda'}\right)
\right\},
\end{equation}
where we have used the notations similar to those of ref.~\cite{kogut}

\begin{equation}
\label{aa}
a_2\equiv\int d^2rr^2\left[A^{-2}({\bf r})B^2({\bf r})-1\right],~
a_3\equiv\int d^2r\left[A^{-2}({\bf r})B^2({\bf r})-1\right].
\end{equation}
Taking into account that $\Lambda'=\Lambda-d\Lambda$ it is straightforward to get

$$
a_2=\alpha_2 K\frac{d\Lambda}{\Lambda^5}\left(1+\frac{\Lambda^2}{m_H^2}\right),~
a_3=\alpha_3 K\frac{d\Lambda}{\Lambda^3}\left(1+\frac{\Lambda^2}{m_H^2}\right).
$$
Here, $\alpha_{2,3}$ stand for some momentum-space-slicing dependent positive constants,
whose concrete values will turn out to be unimportant for the final expressions describing the
RG flow.

Next, since
$a_{2,3}$ occur to be infinitesimal (because they are proportional to $d\Lambda$), the
terms containing these constants on the r.h.s. of eq.~(\ref{ren}) can be treated
in the leading-order approximation of the cumulant expansion that we shall apply for the average over $\psi$. In fact,
we have

$$
\int\limits_{0<p<\Lambda'}^{}{\cal D}\psi({\bf p})\exp\left[
\frac12\int d^2x\psi_{\Lambda'}\left(\partial^2-m_H^2\right)\psi_{\Lambda'}\right]
\exp\left(b\int d^2x {\rm e}^{2\sqrt{K}\psi_{\Lambda'}}f({\bf x})\right)\simeq$$

$$\simeq\exp\left[b{\rm e}^{2KG(0)}\int d^2x f({\bf x})
+\frac{b^2}{2}{\rm e}^{4KG(0)}\int d^2xd^2y\left({\rm e}^{4KG({\bf x}-{\bf y})}-1\right)f({\bf x})f({\bf y})\right],$$
where $f$ is equal either to unity or to $(\partial_\mu\chi_{\Lambda'})^2$, $b\sim a_{2,3}(\xi A(0)B(0))^2$, and

$$G({\bf x})\equiv\int\limits_{0<p<\Lambda'}^{}\frac{d^2p}{(2\pi)^2}
\frac{{\rm e}^{i{\bf p}{\bf x}}}{p^2+m_H^2},~
G(0)=\frac{1}{4\pi}\ln\left(1+\frac{\Lambda'^2}{m_H^2}\right).$$
In order to estimate the parameter of the cumulant expansion,
$\kappa\equiv b{\rm e}^{2KG(0)}\int d^2x\left({\rm e}^{4KG({\bf x})}-1\right)$,
note that we are working in the phase where
monopoles form the plasma, i.e., below the BKT critical temperature $T_c=g^2/2\pi$. By virtue of this fact,
$4K|G({\bf x})|\le 32\pi(\Lambda'/m_H)^2$, that due to the factor $(\Lambda'/m_H)^2$ is generally much smaller than unity.
Owing to that, we get

$$\kappa\simeq b\left(1+\frac{\Lambda'^2}{m_H^2}\right)^{\frac{K}{2\pi}}4K\int d^2x G({\bf x})\simeq
\frac{2bK}{\pi m_H^2}\left[1+\frac{K}{2\pi}\left(\frac{\Lambda'}{m_H}\right)^2\right].$$
Choosing for concreteness $b=a_3(\xi A(0)B(0))^2$ and taking into account that

\begin{equation}
\label{ab}
A(0)\simeq 1-\frac{K}{2}G_h(0)=1-\frac{K}{4\pi}\frac{d\Lambda}{\Lambda},~
B(0)\simeq 1+\frac{K}{2}G_{\phi}(0)\simeq 1+\frac{K}{4\pi}\left(\frac{\Lambda}{m_H}\right)^2
\frac{d\Lambda}{\Lambda},
\end{equation}
we obtain to the leading order:
$\kappa\simeq 2a_3K\xi^2/(\pi m_H^2)$. This quantity possesses the double smallness--
firstly, becase $a_3$ is infinitesimal and secondly, due to the exponential smallness of $\xi$.

Such an extremely rapid convergence of the cumulant expansion then enables us to replace
${\rm e}^{2\sqrt{K}\psi_{\Lambda'}}$ in the terms proportional to $a_{2,3}$ on the r.h.s. of eq.~(\ref{ren})
by the average value of this exponent equal to $\left(1+\frac{\Lambda'^2}{m_H^2}\right)^{\frac{K}{2\pi}}$.
Comparing the so-obtained expression with the initial one, we arrive at the following renormalizations
of the fields and parameters of the Lagrangian:

\begin{equation}
\label{lkm}
\chi_{\Lambda'}^{\rm new}=C\chi_{\Lambda'},~
\psi_{\Lambda'}^{\rm new}=C\psi_{\Lambda'},~
K^{\rm new}=\frac{K}{C^2},~
\mu^{\rm new}=\frac{\mu}{C^2},~
\xi^{\rm new}=A(0)B(0)\xi,
\end{equation}
where

\begin{equation}
\label{rg}
\mu\equiv m_H^2,~ C\equiv\sqrt{1+\frac{Ka_2}{2}(\xi A(0)B(0))^2\left(1+\frac{\Lambda'^2}{m_H^2}\right)^{\frac{K}{2\pi}}}\simeq
\sqrt{1+\frac{Ka_2}{2}(\xi A(0)B(0))^2\left(1+\frac{\Lambda^2}{m_H^2}\frac{K}{2\pi}\right)}.
\end{equation}
Besides that, we obtain the following shift of the free-energy density $F\equiv-\frac{\ln {\cal Z}'}{V}$:

\begin{equation}
\label{endens}
F=F^{\rm new}-a_3(\xi A(0)B(0))^2\left(1+\frac{\Lambda'^2}{m_H^2}\right)^{\frac{K}{2\pi}}\simeq
F^{\rm new}-a_3(\xi A(0)B(0))^2\left(1+\frac{K}{2\pi}\frac{\Lambda'^2}{m_H^2}\right),
\end{equation}
where $V$ is the 2D-volume (i.e., area) of the system.

By making use of the relations~(\ref{ab}), it is further straightforward to derive from eqs.~(\ref{lkm})-(\ref{endens})
the RG equations in the differential form. Those read

$$d\xi=-\frac{K\xi}{4\pi}\left(1-\frac{\Lambda^2}{\mu}\right)\frac{d\Lambda}{\Lambda},~
dK=-\frac{\alpha_2}{2}K^3\xi^2\left[1+\left(\frac{K}{2\pi}+1\right)\frac{\Lambda^2}{\mu}\right]\frac{d\Lambda}{\Lambda^5},$$

$$d\mu=-\frac{\alpha_2}{2}(K\xi)^2\mu\left[1+\left(\frac{K}{2\pi}+1\right)\frac{\Lambda^2}{\mu}\right]\frac{d\Lambda}{\Lambda^5},~
dF=\alpha_3K\xi^2\left[1+\left(\frac{K}{2\pi}+1\right)\frac{\Lambda^2}{\mu}\right]\frac{d\Lambda}{\Lambda^3}.$$
Following the notations of ref.~\cite{kogut}, we shall further make the change of variables from the momentum scale
to the real-space one:
$\Lambda\to a\equiv1/\Lambda$, $d\Lambda\to-d\Lambda$, that obviously modifies the above equations as

\begin{equation}
\label{xi}
d\xi=-\frac{K\xi}{4\pi}\frac{da}{a}\left(1-\frac{1}{\mu a^2}\right),
\end{equation}

\begin{equation}
\label{K}
dK=-\frac{\alpha_2}{2}K^3\xi^2a^3da\left[1+\left(\frac{K}{2\pi}+1\right)\frac{1}{\mu a^2}\right],
\end{equation}

\begin{equation}
\label{mu}
d\mu=-\frac{\alpha_2}{2}(K\xi)^2\mu a^3da\left[1+\left(\frac{K}{2\pi}+1\right)\frac{1}{\mu a^2}\right],
\end{equation}

\begin{equation}
\label{F}
dF=\alpha_3K\xi^2ada\left[1+\left(\frac{K}{2\pi}+1\right)\frac{1}{\mu a^2}\right].
\end{equation}

Our main aim below is to derive from eqs.~(\ref{xi})-(\ref{mu}) the leading-order corrections in $(\mu a^2)^{-1}$
to the BKT RG flow in the vicinity of the critical point. This point is known to be~\cite{bkt, rev, kogut}
$K^{(0)}_{\rm cr.}=8\pi$, $y^{(0)}_{\rm cr.}=0$, where $y\equiv \xi a^2$, and the superscription ``${\,}{}^{(0)}{\,}$'' denotes
the zeroth order in the $(\mu a^2)^{-1}$-expansion. These values of $K^{(0)}_{\rm cr.}$ and $y^{(0)}_{\rm cr.}$
will be recovered below. Besides that, it will be
demonstrated that in the critical region, $\mu$ is evolving very weakly. Owing to this fact, the initial assumption
on the largeness of $\mu$ (namely, that it is of the order of $m_W^2$),
will be preserved by the RG flow, at least in that region.
This enables us to treat $\mu$ almost as a constant and seek for the corrections to the RG flow of $K^{(0)}$
in powers of $(\mu a^2)^{-1}$. The zeroth-order equation stemming from eq.~(\ref{K}) then reads

\begin{equation}
\label{K0}
dK^{(0)}=-\frac{\alpha_2}{2}K^{(0){\,}3}\xi^2a^3da.
\end{equation}
Respectively, the zeroth-order in $(\mu a^2)^{-1}$ equation for $y$ has the form

\begin{equation}
\label{y0}
dy^{(0){\,}2}=2\frac{da}{a}y^{(0){\,}2}x,
\end{equation}
where $x\equiv 2-\frac{K^{(0)}}{4\pi}$.
Equations~(\ref{K0}) and (\ref{y0}) yield
the above-mentioned leading critical value of $K$, $K^{(0)}_{\rm cr.}$. Note that the respective critical temperature following
from this value is equal to $T_c$~\footnote{This result coincides with the value of $T_c$ obtained in ref.~\cite{nk} upon the evaluation
of the mean squared separation in the monopole-antimonopole molecule at high temperatures, in the compact-QED limit (see the discussion
below).}. Next, with this value of $K^{(0)}_{\rm cr.}$,
eq.~(\ref{K0}) can be rewritten in the vicinity of the critical point as

\begin{equation}
\label{x}
dx=(8\pi)^2\alpha_2\xi^2a^3da.
\end{equation}
Introducing further instead of $y$ the new variable $z=(8\pi)^2\alpha_2y^2$ and performing the rescaling
$a^{\rm new}=a\sqrt{8\pi\alpha_2/\alpha_3}$
we get from eqs.~(\ref{F}), (\ref{y0}), and~(\ref{x}) the following system of equations:

\begin{equation}
\label{star}
dz^{(0)}=2\frac{da}{a}xz^{(0)},~
dx=z^{(0)}\frac{da}{a},~
dF^{(0)}=z^{(0)}\frac{da}{a^3}.
\end{equation}
These equations yield the standard RG flow in the vicinity of the critical point, $x=z^{(0)}=0$,
which reads~\cite{rev, kogut} $z^{(0)}-x^2=\tau$, where
$\tau\propto(T_c-T)/T_c$ is some constant. In particular, $x\simeq\sqrt{z}$ at $T\to T_c-0$. Owing to the
first of eqs.~(\ref{star}) this relation yields
$\left(z_{\rm in.}^{(0)}\right)^{-1/2}-\left(z^{(0)}\right)^{-1/2}=\ln\left(a/a_{\rm in.}\right)$, where the superscription ``${\,}{}_{\rm in.}{\,}$''
means the initial value.
Taking into account that $z_{\rm in.}^{(0)}$ is exponentially small, while
$z^{(0)}\sim 1$ (the value at which the growth of $z^{(0)}$ stops), we obtain in the case $x_{\rm in.}\le\sqrt{\tau}$:
$\ln\left(a/a_{\rm in.}\right)\sim\left(z_{\rm in.}^{(0)}\right)^{-1/2}\sim\tau^{-1/2}$. According to this relation,
at $T\to T_c-0$, the correlation radius diverges with an essential singularity as
$a(\tau)\sim\exp\left({\rm const}/\sqrt{\tau}\right)$. (In the molecular phase, the correlation
radius becomes infinite due to the short-rangeness of the molecular fields.)
As far as the leading part of the free-energy density is concerned, it scales as $F^{(0)}\sim a^{-2}$ and therefore
remains continuous in the critical region. Moreover, the correction to this behaviour stemming from the finiteness
of the Higgs-boson mass [the last term on the r.h.s. of eq.~(\ref{F})] is clearly of the same functional form,
$\sim\exp\left(-{\rm const}'/\sqrt{\tau}\right)$, i.e., it is also continuous.

We are now in the position to address the leading-order [in $(\mu a^2)^{-1}$] corrections to the
above-discussed BKT RG flow of $K^{(0)}$ and $z^{(0)}$. To this end, let us represent $K$ and $z$ as
$K=K^{(0)}+K^{(1)}/(\mu a^2)$, $z=z^{(0)}+z^{(1)}/(\mu a^2)$,
that by virtue of eqs.~(\ref{xi}) and (\ref{K}) leads to the following novel equations:

\begin{equation}
\label{K1}
dK^{(1)}-2K^{(1)}\frac{da}{a}=-4\pi\frac{da}{a}\left(z^{(0)}+\frac{z^{(1)}}{\mu a^2}\right)\left(1+\frac{K^{(0)}}{2\pi}+
\frac{3K^{(1)}}{K^{(0)}}\right),
\end{equation}

\begin{equation}
\label{z1}
dz^{(1)}-2z^{(1)}\frac{da}{a}=-2\frac{da}{a}\left[z^{(1)}\left(\frac{K^{(0)}}{4\pi}-2\right)+\frac{z^{(0)}}{4\pi}
\left(K^{(1)}-K^{(0)}\right)\right].
\end{equation}
In the vicinity of the critical point, we can
insert into eq.~(\ref{z1}) the above-obtained critical values of $K^{(0)}$ and $z^{(0)}$, that yields:

\begin{equation}
\label{z11}
dz^{(1)}=2z^{(1)}\frac{da}{a}.
\end{equation}
Therefore, $z^{(1)}=C_1a^2$, where $C_1$ is the integration constant of dimensionality $({\rm mass})^2$, $C_1\ll\mu$. Inserting further
this solution into eq.~(\ref{K1}), considered in the vicinity of the critical point, we obtain the following equation:

\begin{equation}
\label{K11}
dK^{(1)}-2K^{(1)}\frac{da}{a}=-\frac{4\pi C_1}{\mu}\frac{da}{a}\left(\frac{3K^{(1)}}{8\pi}+5\right).
\end{equation}
Its integration is straightforward and yields

\begin{equation}
\label{C2}
K^{(1)}=C_2\left(\mu a^2\right)^{1-\frac{3C_1}{4\mu}}+\frac{40\pi C_1}{4\mu-3C_1},
\end{equation}
where the dimensionless integration constant $C_2$ should be much smaller than $(\mu a^2)^{\frac{3C_1}{4\mu}}$. [Note that the last addendum
in eq.~(\ref{C2}) is positive.] Therefore, the total correction, $K^{(1)}/(\mu a^2)$, approximately scales with $a$ in the critical region as
$\frac{40\pi C_1}{(4\mu-3C_1)\mu a^2}$. We see that at the critical point, this expression
vanishes due to the divergence of the correlation radius.
Note also that eq.~(\ref{C2}) can obviously be rewritten as the following dependence of $K^{(1)}/(\mu a^2)$ on $z^{(1)}$
($z^{(1)}\ll\mu a^2$):

$$
\frac{K^{(1)}}{\mu a^2}=C_2\left(\mu a^2\right)^{-\frac{3z^{(1)}}{4\mu a^2}}+
\frac{40\pi z^{(1)}}{\left(4\mu a^2-3z^{(1)}\right)\mu a^2}.
$$
With the above-discussed critical behaviour of the correlation radius, $a(\tau)$, this relation determines the correction to the
BKT RG flow, $z^{(0)}-\left(2-\frac{K^{(0)}}{4\pi}\right)^2=\tau$.

Clearly, the critical values of $z^{(1)}$ and $K^{(1)}$ determined by the
fixed point of eqs.~(\ref{z11}) and~(\ref{K1}) [or~(\ref{K11})]
read $z^{(1)}_{\rm cr.}=K^{(1)}_{\rm cr.}=0$. This means
that the finiteness of the Higgs-boson mass does not change the value of the BKT critical temperature, $T_c$.
Note that this fact can also be seen by the evaluation of the
mean squared separation in the monopole-antimonopole molecule. The expression for this quantity stems
from the statistical weight in the monopole ensemble, which at finite temperature reads [cf. eq.~(\ref{pf})]

$$
\exp\left[\frac{K}{4\pi}
\sum\limits_{{a,b=1\atop a\ne b}}^{N}\left(q_aq_b\ln\left(\bar\mu\left|{\bf z}_a-{\bf z}_b\right|\right)+
K_0\left(m_H\left|{\bf z}_a-{\bf z}_b\right|\right)\right)\right].$$
Here, $\bar\mu$ stands for the IR cutoff, $K_0$ is the modified Bessel function, and ${\bf z}_{a,b}$ are the 2D-vectors. The desired
mean squared separation is then given by the following formula:

$$
\left<L^2\right>=\frac{\int\limits_{|{\bf x}|>m_W^{-1}}^{} d^2{\bf x}|{\bf x}|^{2-\frac{8\pi T}{g^2}}\exp\left[\frac{4\pi T}{g^2}K_0\left(m_H|{\bf x}|
\right)\right]}{\int\limits_{|{\bf x}|>m_W^{-1}}^{} d^2{\bf x}|{\bf x}|^{-\frac{8\pi T}{g^2}}\exp\left[\frac{4\pi T}{g^2}K_0\left(m_H|{\bf x}|
\right)\right]}.$$
In the case $m_H\sim m_W$ under study, the exponential factors in the numerator and denominator of this equation can be disregarded,
and we obtain
$\left<L^2\right>\simeq\frac{4\pi T-g^2}{2m_W^2\left(2\pi T-g^2\right)}$,
that yields the value of the critical temperature equal to $T_c$. Besides that, it is straightforward to see that in the weak-coupling
regime under study, the value of $\sqrt{\left<L^2\right>}$ is exponentially smaller than the characteristic distance in the
monopole plasma, $\zeta^{-1/3}$, i.e., molecules are very small-sized with respect to that distance.

Finally, in order to justify the above-adapted approximation under which $\mu$ was treated as a constant,
we should check that under the RG flow, it is really evolving only weakly. To this end, let us pass in
eq.~(\ref{mu}), considered in the critical region, from the variable $\xi$ to the above-introduced variable $z$
and perform again the rescaling
$a\to a^{\rm new}$. This yields $\frac{d\mu}{\mu}=-\frac{z}{2}\frac{da}{a}$ or $d\mu=-\frac{C_1}{2}\frac{da}{a}$.
Since $C_1\ll\mu$, we conclude that $\frac{|d\mu|}{\mu}\ll\frac{da}{a}$. This inequality means that in the vicinity of the
BKT critical point, $\mu$ is really evolving weakly. This justifies its treatment as a large (with respect to $\Lambda^2$)
constant quantity, approximately equal to its initial value of the order of $m_W^2$.

\section{SU(N)-case.}

The $SU(N)$-generalization of the action~(\ref{1}), stemming from the $SU(N)$
Georgi-Glashow model, has the form

\begin{equation}
\label{s}
S=\int d^3x\left[\frac12(\partial_\mu\vec\chi)^2+\frac12(\partial_\mu\psi)^2+
\frac{m_H^2}{2}\psi^2-2\zeta{\rm e}^{g_m\psi}\sum\limits_{i}^{}
\cos\left(g_m\vec q_i\vec\chi\right)\right].
\end{equation}
Here, $\sum\limits_{i}^{}\equiv\sum\limits_{i=1}^{N(N-1)/2}$, and
$\vec q_i$'s are the positive root vectors of the group $SU(N)$.
As well as the field $\vec\chi$, these vectors are $(N-1)$-dimensional.
Note that the $SU(3)$- and the general $SU(N)$-versions of the action~(\ref{s}), thus incorporating
the effects of the Higgs field, have been discussed
in refs.~\cite{nd, mpla}. The compact-QED limit of the $SU(N)$-case
has been studied (at zero temperature) in refs.~\cite{wd} and~\cite{sn}.
Here, similarly to all the above-mentioned papers, we have assumed that W-bosons corresponding to
different root vectors have the same masses.

Considering the action~(\ref{s}) at finite temperature and applying to the resulting
dimensionally-reduced theory the RG procedure of the previous Section, we arrive at the following analogue
of eq.~(\ref{ren}):

$$
{\cal Z}_\Lambda=\int\limits_{0<p<\Lambda'}^{}{\cal D}\vec\chi({\bf p}){\cal D}\psi({\bf p})\exp\left\{
\frac12\int d^2x\psi_{\Lambda'}\left(\partial^2-m_H^2\right)\psi_{\Lambda'}+(\xi A(0)B(0))^2\int d^2x{\rm e}^{2\sqrt{K}\psi_{\Lambda'}}
\sum\limits_{ij}^{}a_3^{ij}\times\right.$$

$$\times\cos\left[\sqrt{K}\left(\vec q_i-\vec q_j\right)\vec\chi_{\Lambda'}\right]+
2\xi A(0)B(0)\int d^2x{\rm e}^{\sqrt{K}\psi_{\Lambda'}}\sum\limits_{i}^{}\cos\left(\sqrt{K}\vec q_i\vec\chi_{\Lambda'}\right)-
\frac12\int d^2x\Biggl[\delta^{ab}+$$

\begin{equation}
\label{ren1}
\left.+(\xi A(0)B(0))^2
\frac{K}{8}{\rm e}^{2\sqrt{K}\psi_{\Lambda'}}\sum\limits_{ij}^{}a_2^{ij}\left(\vec q_i+\vec q_j\right)^{\alpha}
\left(\vec q_i+\vec q_j\right)^{\beta}
\cos\left[\sqrt{K}\left(\vec q_i-\vec q_j\right)\vec\chi_{\Lambda'}\right]\Biggr]\Bigl(\partial_\mu\chi_{\Lambda'}^{\alpha}\Bigr)
\Bigl(\partial_\mu\chi_{\Lambda'}^{\beta}\Bigr)\right\}.
\end{equation}
Here, $\alpha,\beta=1,\ldots,(N-1)$, and we have introduced the notations similar to (\ref{aa}),

$$a_2^{ij}\equiv\int d^2rr^2\left[B^2({\bf r}){\rm e}^{K\vec q_i\vec q_jG_h({\bf r})}-1\right],~
a_3^{ij}\equiv\int d^2r\left[B^2({\bf r}){\rm e}^{K\vec q_i\vec q_jG_h({\bf r})}-1\right],$$
so that at $\Lambda'=\Lambda-d\Lambda$,

$$a_2^{ij}=\alpha_2 K\frac{d\Lambda}{\Lambda^5}\left(\vec q_i\vec q_j+\frac{\Lambda^2}{m_H^2}\right),~
a_3^{ij}=\alpha_3 K\frac{d\Lambda}{\Lambda^3}\left(\vec q_i\vec q_j+\frac{\Lambda^2}{m_H^2}\right).$$
The main difference of eq.~(\ref{ren1}) from eq.~(\ref{ren}) is due to the terms containing
$\cos\left[\sqrt{K}\left(\vec q_i-\vec q_j\right)\vec\chi_{\Lambda'}\right]$, that violate the
RG invariance. Nevertheless, this invariance approximately holds, since the respective sums are dominated
by the terms with $i=j$. Working within this approximation and making use of the identity~\cite{group, mpla} $\sum\limits_{i}^{}
q_i^\alpha q_i^\beta=\frac{N}{2}\delta^{\alpha\beta}$, we obtain

$$
{\cal Z}_\Lambda\simeq\int\limits_{0<p<\Lambda'}^{}{\cal D}\vec\chi({\bf p}){\cal D}\psi({\bf p})\exp\left\{
\frac12\int d^2x\psi_{\Lambda'}\left(\partial^2-m_H^2\right)\psi_{\Lambda'}+\right.$$

$$+a_3\frac{N(N-1)}{2}(\xi A(0)B(0))^2\left(
1+\frac{\Lambda'^{2}}{m_H^2}\right)^{\frac{K}{2\pi}}V+
2\xi A(0)B(0)\int d^2x{\rm e}^{\sqrt{K}\psi_{\Lambda'}}\sum\limits_{i}^{}\cos\left(\sqrt{K}\vec q_i\vec\chi_{\Lambda'}\right)-$$

\begin{equation}
\label{double}
\left.-\frac12\int d^2x\Biggl[1+\left(
1+\frac{\Lambda'^{2}}{m_H^2}\right)^{\frac{K}{2\pi}}\frac{NKa_2}{4}(\xi A(0)B(0))^2\Biggr]\left(\partial_\mu\vec\chi_{\Lambda'}
\right)^2\right\}.
\end{equation}
This expression has again been derived in the leading order of the cumulant
expansion applied in the course of the average over $\psi$.

The shift of the free-energy density stemming from eq.~(\ref{double}) [cf. eq.~(\ref{endens})] reads

\begin{equation}
\label{en1}
F^{\rm new}-F\simeq
a_3\frac{N(N-1)}{2}(\xi A(0)B(0))^2\left(
1+\frac{K}{2\pi}\frac{\Lambda'^{2}}{m_H^2}\right).
\end{equation}
As far as the renormalization of fields and coupling constants is concerned, it is given by eq.~(\ref{lkm}), where
the first equation should be modified as $\vec\chi_{\Lambda'}{}^{\rm new}=C\vec\chi_{\Lambda'}$, and
the parameter $C$ from eq.~(\ref{rg}) now reads

\begin{equation}
\label{CN}
C=\sqrt{1+\frac{KNa_2}{4}(\xi A(0)B(0))^2\left(1+\frac{\Lambda'^2}{m_H^2}\right)^{\frac{K}{2\pi}}}\simeq
\sqrt{1+\frac{KNa_2}{4}(\xi A(0)B(0))^2\left(1+\frac{\Lambda^2}{m_H^2}\frac{K}{2\pi}\right)}.
\end{equation}
From eqs.~(\ref{en1}) and (\ref{CN}) we deduce that in the $SU(N)$-case
the RG flow of couplings and of the free-energy density is identical to that
of the $SU(2)$-case. Indeed, all the $N$-dependence can be absorbed into the constants $\alpha_{2,3}$
by rescaling them as $\bar\alpha_2\equiv N\alpha_2/2$, $\bar\alpha_3=N(N-1)\alpha_3/2$ and further redefining
[cf. the notations introduced after eq.~(\ref{x})]
$\bar z=(8\pi)^2\bar\alpha_2y^2$ and $\bar a^{\rm new}=a\sqrt{8\pi\bar\alpha_2/\bar\alpha_3}$.
In particular, the critical temperature $T_c$
remains the same as in the $SU(2)$-case. (This follows also from the estimate of the mean squared separation in the monopole-antimonopole
molecule, if one takes into account that the square of any root vector is equal to unity.)
Thus, the principal difference of the $SU(N)$-case, $N>2$, from the
$SU(2)$-one is that while in the $SU(2)$-case
the RG invariance is exact (modulo the negligibly small higher-order terms of the cumulant expansion applied
to the average over $\psi$), in the $SU(N)$-case it is only approximate, even in the compact-QED limit of the model.

\section{Conclusions.}
In the present letter, we have explored the influence of the Higgs field to the RG flow in the finite-temperature
3D Georgi-Glashow model and in its $SU(N)$-generalization. In this investigation, the effects of W-bosons have been
disregarded, and the analysed RG flow describes the BKT phase transition in the monopole ensemble (rather than the real
deconfining phase transition, for whose dynamics W-bosons play the crucial r\^ole).
Further, the Higgs-field mass, $m_H$, was supposed to be
large, namely of the order of the W-boson one, but not infinite, as it takes place in the compact-QED limit of the model.
There have been derived
the leading $(\Lambda/m_H)$-corrections (where $\Lambda$ is the momentum scale we are working at)
to the standard BKT RG flow in the vicinity of the critical point. In particular, it turned out that
the correction to the magnetic coupling constant
of the dimensionally-reduced theory is inversly proportional to the second power of the correlation radius
and therefore vanishes at the critical point.
Besides that, there has been derived the RG equation describing the evolution of $m_H$ itself, which shows
that $m_H$ evolves so weakly in the vicinity of the critical point that it can be treated as a constant
with a high accuracy. It also turned out that the derived corrections to the RG flow do not affect the position
of the critical point. In particular, the value of the BKT critical temperature remains unchanged if one accounts for
the finiteness of the Higgs-boson mass. The RG equation describing the evolution of the free-energy density with the account for the
Higgs-inspired correction has also been derived. According to this equation, the free-energy density remains continuous
in the critical region, and the above-mentioned correction does not violate this property.
It has further been demonstrated that contrary to the $SU(2)$-case, in the
$SU(N)$-model at $N>2$, the RG invariance holds only approximately, even in the compact-QED limit. Namely, it holds modulo the approximation
$\sum\limits_{ij}^{}f_{ij}\cos\left[\left(\vec q_i-\vec q_j\right)\vec\chi\right]\simeq\sum\limits_{i}^{}f_{ii}$,
where $\vec q_i$'s are the positive root vectors of the group $SU(N)$. Within this approximation, the RG flow (and consequently,
the critical temperature) in the $SU(N)$-model
is identical to the one of the $SU(2)$-case,
since all the $N$-dependence can then be removed upon the appropriate rescaling.

\section{Acknowledgments.}
The author is grateful for useful discussions
to Profs. A.~Di~Giacomo and Yu.M.~Makeenko, and to Dr. N.O.~Agasian. He is also grateful to
Prof. A.~Di~Giacomo and to the whole staff of the Physics Department of the
University of Pisa for cordial hospitality.
The work has been supported by INFN and partially by
the INTAS grant Open Call 2000, Project No. 110.


\begin{thebibliography}{100}
\bibitem{1}A.M. Polyakov, Nucl. Phys. {\bf B 120} (1977) 429.
%%CITATION = NUPHA,B120,429;%%


\bibitem{nd}
N.O. Agasian and D. Antonov, JHEP {\bf 06} (2001) 058.
%%CITATION = JHEPA,0106,058;%%

\bibitem{mpla}
D. Antonov, Mod. Phys. Lett. {\bf A 17} (2002) 279.
%%CITATION = HEP-TH 0201013;%%


\bibitem{bps}
M.K. Prasad and C.M. Sommerfield, Phys. Rev. Lett. {\bf 35} (1975) 760;
E.B. Bogomolny, Sov. J. Nucl. Phys. {\bf 24} (1976) 449.
%%CITATION = PRLTA,35,760;%%
%%CITATION = YAFIA,24,861;%%


\bibitem{dietz}
K. Dietz and Th. Filk, Nucl. Phys. {\bf B 164} (1980) 536.
%%CITATION = NUPHA,B164,536;%%


\bibitem{kirk}
T.W. Kirkman and C.K. Zachos, Phys. Rev. {\bf D 24} (1981) 999.
%%CITATION = PHRVA,D24,999;%%


\bibitem{bkt}
V.L. Berezinsky, Sov. Phys.- JETP {\bf 32} (1971) 493;
J.M. Kosterlitz and D.J. Thouless, J. Phys. {\bf C 6}
(1973) 1181; J.M. Kosterlitz, J. Phys. {\bf C 7} (1974) 1046.
%%CITATION = JTPHE,32,493;%%
%%CITATION = JPCBA,6,1181;%%
%%CITATION = JPCBA,7,1046;%%


\bibitem{rev}
J. Zinn-Justin,
{\it Quantum Field Theory and Critical Phenomena} (Oxford Univ. Press,
2nd edn., New York, 1993).



\bibitem{W}
G. Dunne, I.I. Kogan, A. Kovner, and B. Tekin, JHEP {\bf 01} (2001) 032;
I.I. Kogan, A. Kovner, and B. Tekin,
JHEP {\bf 03} (2001) 021;
Phys. Rev. {\bf D 63} (2001) 116007;
JHEP {\bf 05} (2001) 062;
I.I. Kogan, A. Kovner, and M. Schvellinger,
JHEP {\bf 07} (2001) 019.
%%CITATION = JHEPA,0101,032;%%
%%CITATION = JHEPA,0103,021;%%
%%CITATION = JHEPA,0105,062;%%
%%CITATION = JHEPA,0107,019;%%
%%CITATION = PHRVA,D63,116007;%%


\bibitem{PLB}
D. Antonov, {\tt hep-th/0204114} (to appear in Phys. Lett. {\bf B}).
%%CITATION = HEP-TH 0204114;%%

\bibitem{nk}
N.O. Agasyan and K. Zarembo, Phys. Rev. {\bf D 57} (1998) 2475.
%%CITATION = PHRVA,D57,2475;%%

\bibitem{kogut}
J.B. Kogut, Rev. Mod. Phys. {\bf 51} (1979) 659;
B. Svetitsky and L.G. Yaffe, Nucl. Phys. {\bf B 210 [FS6]} (1982) 423.
%%CITATION = RMPHA,51,659;%%
%%CITATION = NUPHA,B210,423;%%


\bibitem{wd}
S.R. Wadia and S.R. Das, Phys. Lett. {\bf B 106} (1981) 386; Erratum-ibid.
{\bf B 108} (1982) 435.
%%CITATION = PHLTA,B106,386;%%


\bibitem{sn}
N.J. Snyderman, Nucl. Phys. {\bf B 218} (1983) 381.
%%CITATION = NUPHA,B218,381;%%


\bibitem{group}
R. Gilmore, {\it Lie groups, Lie algebras, and some of their applications}
(J. Wiley \& Sons, New York, 1974).



\end{thebibliography}
\end{document}